\documentclass[aps,prl,reprint,showpacs,groupedaddress]{revtex4-1}

\usepackage{amsmath}
\usepackage{amssymb}
\usepackage{graphicx}
\usepackage[capitalize]{cleveref}
\usepackage{color}

\def\pl{\partial}

\def\al{\alpha}
\def\bt{\beta}
\def\Ga{\Gamma}
\def\ga{\gamma}
\def\de{\delta}

\def\te{\theta}

\def\lam{\lambda}

\def\ep{\epsilon}

\def\l{\left (}
\def\r{\right )}

\def\fr{\frac}
\def\la{\label}

\def\vs{\vspace}

\def\ran{\rangle}
\def\lan{\langle}

\def\tl{\tilde}
\def\tm{\times}

\begin{document}

\newcommand{\ba}[1]{\begin{array}{#1}} \newcommand{\ea}{\end{array}}



\def\Journal#1#2#3#4{{#1} {\bf #2}, #3 (#4)}

\def\NCA{\em Nuovo Cimento}
\def\NIM{\em Nucl. Instrum. Methods}
\def\NIMA{{\em Nucl. Instrum. Methods} A}
\def\NPB{{\em Nucl. Phys.} B}
\def\PRL{\em Phys. Rev. Lett.}
\def\PRD{{\em Phys. Rev.} D}
\def\ZPC{{\em Z. Phys.} C}

\def\st{\scriptstyle}
\def\sst{\scriptscriptstyle}
\def\mco{\multicolumn}
\def\epp{\epsilon^{\prime}}
\def\vep{\varepsilon}
\def\ra{\rightarrow}
\def\ppg{\pi^+\pi^-\gamma}
\def\vp{{\bf p}}
\def\ko{K^0}
\def\kb{\bar{K^0}}
\def\al{\alpha}
\def\ab{\bar{\alpha}}

\def\np{Nucl. Phys. {\bf B}}
\def\mpl{Mod. Phys. {\bf A}}\def\ijmp{Int. J. Mod. Phys. {\bf A}}
\def\cmp{Comm. Math. Phys.}\def\prd{Phys. Rev. {\bf D}}

\def\oa{\bigcirc\!\!\!\! a}
\def\ob{\bigcirc\!\!\!\! b}
\def\oc{\bigcirc\!\!\!\! c}
\def\oi{\bigcirc\!\!\!\! i}
\def\oj{\bigcirc\!\!\!\! j}
\def\ok{\bigcirc\!\!\!\! k}
\def\ve{\vec e}\def\vk{\vec k}\def\vn{\vec n}\def\vp{\vec p}
\def\vv{\vec v}\def\vx{\vec x}\def\vy{\vec y}\def\vz{\vec z}

\newcommand{\AdS}{\mathrm{AdS}}
\newcommand{\dd}{\mathrm{d}}
\newcommand{\eee}{\mathrm{e}}
\newcommand{\sgn}{\mathop{\mathrm{sgn}}}

\def\a{\alpha}
\def\b{\beta}
\def\g{\gamma}

\newcommand\lsim{\mathrel{\rlap{\lower4pt\hbox{\hskip1pt$\sim$}}
    \raise1pt\hbox{$<$}}}
\newcommand\gsim{\mathrel{\rlap{\lower4pt\hbox{\hskip1pt$\sim$}}
    \raise1pt\hbox{$>$}}}

\newcommand{\beq}{\begin{equation}}
\newcommand{\eeq}{\end{equation}}
\newcommand{\bea}{\begin{eqnarray}}
\newcommand{\eea}{\end{eqnarray}}
\newcommand{\noi}{\noindent}


\title{Chaotic Inflation from the MSSM Along Flat $D$-Term Trajectory}

\author{Zurab Tavartkiladze}
\email[]{zurab.tavartkiladze@gmail.com}

\affiliation{Center for Elementary Particle Physics, ITP, Ilia State University, 0162 Tbilisi, Georgia}


\begin{abstract}
Within the MSSM we propose the chaotic inflationary scenario in which
 the inflaton field is a combination of sleptons and the Higgs field states evolving
along the $D$-term flat direction.
 In the inflation and postinflation reheating processes, a decisive role is played by the MSSM Yukawa
   superpotential. The vacuum energy during the inflationary era is mainly from the muonic Yukawa coupling,
while the inflaton decay and subsequent reheating process dominantly proceeds due to the strange quark
Yukawa term. Because of these,  the presented scenario is predictive and the results obtained  agree well with cosmological
observations. In particular, the scalar spectral index and the tensor-to-scalar ratio
are respectively, $n_s\simeq 0.966$ and $r=0.00117$.
The reheating temperature is found to be $T_r\simeq 7.2\tm 10^7$~GeV.
\end{abstract}



\maketitle

\section{I. Introduction}
\vs{-0.3cm}
Recent cosmological data released by the Planck Collaboration \cite{Aghanim:2018eyx, Akrami:2018odb} provide a great opportunity to probe theoretical inflationary models with improved accuracy. Upon model building, there has been 
much effort made to find a close connection between models of particle physics and inflation. In spite of this, to our knowledge, there are no models with predicted observables expressed in terms of the Standard Model or MSSM (the minimal supersymmetric Standard Model) parameters. In this work, aiming to fill this gap, within the MSSM framework we propose a model of inflation in which predictions are made in terms of MSSM superpotential couplings.

Many problems of the big bang cosmology are solved by inflation \cite{first-infl} - the era
of the exponential expansion of the Universe driven by the inflaton field.
In order to have successful inflation, inflaton's slow roll should be ensured (perhaps by some symmetry).
 For this purpose the supersymmetric (SUSY) constructions are very efficient \cite{susy-infl}.
  The MSSM   is highly motivated from the phenomenological and theoretical viewpoints.
However, upon building the  SUSY inflation models, in most cases, one goes beyond the MSSM and introduces additional singlet superfield(s). Such extensions, involving additional parameters,  increase the degree of arbitrariness.
 Since the MSSM involves  the sleptons and squarks - the superpartners of the SM matter - there is a
 loophole to use either one or a combination of these states as an inflaton.
  The possible role of the slepton and/or
 squark condensates, along the flat directions, for baryogenesis  has been emphasized in  \cite{Affleck:1984fy}
 and investigated further  in \cite{Dine:1995kz}.
 In exploiting  sleptons and/or squarks, for inflation, numerous works have been proposed \cite{Allahverdi:2006iq}.
 However, these models involve many new parameters due to higher order operators (which are needed
to realize desirable inflation).

In this paper we propose a scenario based on the MSSM (without its extension).
We show that field configuration along the $D$-flat direction, but with a nonvanishing $F$-term(s),
can give very successful and predictive inflationary scenario. Along the MSSM Yukawa superpotential
couplings, allowing (as we show) us to have  successful inflation, we
use the nonminimal  K\" ahler potential of the form
\begin{eqnarray}
\la{totalK}
&{\cal K}=-M_{Pl}^2\ln (1-\fr{1}{M_{Pl}^2}\sum_{I}\Phi_I^\dag e^{-V}\Phi_I) 
\end{eqnarray}
(under $\Phi_I$ the MSSM chiral superfields are assumed),
which in the small field limit ($\Phi_I\!\ll \!M_{Pl}$) gets the canonical form ${\cal K}\to \sum_{I}\Phi_I^\dag e^{-V}\Phi_I$.
The structure of (\ref{totalK}), together with the MSSM supeppotential, turns out to be crucial for successful inflation \cite{logK},
 with observables determined in terms of the MSSM parameters.

\section{II. The Choice of $D$-term Flat Direction}
\vs{-0.3cm}

The MSSM superpotential involves the Yukawa part $W_Y$ and the $\mu $-term
\begin{eqnarray}
\label{WMSSM}
&W_{MSSM}=W_Y+\mu h_uh_d~,~~~~
\nonumber \\
&{\rm with}~~~W_Y= e^cY_Elh_d+ qY_Dd^ch_d+ qY_Uu^ch_u~.
\end{eqnarray}
Without loss of any generality, we choose the field basis such that
the Yukawa matrices are:
\begin{eqnarray}
\label{Yuk-basis}
&Y_E=Y_E^{\rm Diag}={\rm Diag}\l \lam_e, \lam_{\mu }, \lam_{\tau }\r ,
\nonumber \\
&Y_D=Y_D^{\rm Diag} ,~Y_U=V_{CKM}^TY_U^{\rm Diag}.
\end{eqnarray}

Moreover, the MSSM  $D$-terms, corresponding to the $U(1)_Y$, $SU(2)_w$ and $SU(3)_c$, are respectively,
\begin{eqnarray}
\label{D-terms}
&D_Y=|h_d|^2-|h_u|^2-2|\tl e^c_{\al}|^2+|\tl l_{\al}|^2
\nonumber \\
&-\fr{1}{3}|\tl q_{\al}|^2+\fr{4}{3}|\tl u^c_{\al}|^2-
\fr{2}{3}|\tl d^c_{\al}|^2 ~,
\nonumber \\
&D^i_{SU(2)}=\!\fr{1}{2}\!\l h_d^\dag \tau^ih_d-h_u^\dag \tau^ih_u+\tl{l}_{\al }^{\dag } \tau^i\tl l_{\al}
+ \tl{q}_{\al }^{\dag } \tau^i\tl q_{\al}\r ,
\nonumber \\
&D^a_{SU(3)}=\!\fr{1}{2}\!\l \tl{q}_{\al }^{\dag } \lam^a\tl q_{\al}-\tl{u}_{\al }^{c\dag } \lam^a\tl u^c_{\al}-
\tl{d}_{\al }^{c\dag } \lam^a\tl d^c_{\al} q_{\al}\r ,
\end{eqnarray}
 In (\ref{D-terms}) the summation under  the $\al=1,2,3$ family index is assumed. Here, $\tau^i/2$ and $\lam^a/2$ stand for
$SU(2)_w$ and $SU(3)_c$ generators, respectively ($i=1,2,3$, $a=1,\cdots ,8$). While there are numerous solutions of the
$D$-term  flat directions, classified in \cite{Gherghetta:1995dv}, we choose the one involving the slepton states
($e^cll$-type flat direction) and the scalar component of $h_d$.
 In particular, we consider the VEV configuration
\beq
\lan \tl e^c_1\ran =z,~
 \lan \tl l_2\ran =
\left(
  \begin{array}{c}
    0 \\
    z \\
  \end{array}
  \right) ,~
  \lan \tl l_3\ran =
\left(
  \begin{array}{c}
    cz \\
    0\\
  \end{array}
\right),
\lan h_d\ran =
\left(
  \begin{array}{c}
    sz \\
    0\\
  \end{array}
\right),
\la{VEVs}
\eeq
where the definitions $\cos \te \equiv c$ and $\sin \te \equiv s$ are introduced, with $\te $ determined later.
One can readily check that with the (\ref{VEVs}) configuration (and with zero VEVs of all remaining fields), for arbitrary values of $z$ and $\te $, the $D$-terms  of $U(1)_Y$, $SU(2)_w$ and $SU(3)_c$ [given in (\ref{D-terms})] vanish.
As will be shown in the next section, this choice leads to successful inflation.

As far as the superpotential couplings are concerned, from (\ref{WMSSM}) we can derive nonvanishing $F$-terms, which are given by
\beq
F_{e^{-}}^* =\lam_esz^2 ~,~~ F_{\mu^c}^* =\lam_{\mu }sz^2 ~,
\la{Fterms}
\eeq
where we have ignored the $\mu $-term, which is too small ($\sim $few TeV) and is irrelevant for our studies.
The  numerical value of the factor $s$ appearing in (\ref{Fterms})
(which turns out to be $s\!\ll \!1$) will be determined  from $A_s$ - the amplitude of the curvature perturbations.
On the other hand, there is a possibility to fix $s$ from the theory. For instance, with additional superpotential couplings
$\lam_1(e^c_3l_2l_3)(e^c_1l_2l_3)-\lam_2(e^c_3l_2h_d)(e^c_1l_2h_d)$, and from the $F_{e^c_3}=0$ condition, one gets
$s\!\simeq \!\!\sqrt{\lam_1/\lam_2}$. Suitable selection of the $\lam_{1,2}$ couplings will give $s$'s desirable values. At the same time, such high dimensional terms (with $\lam_{1,2}$ couplings), being fully consistent, have no impact on low energy phenomenology \cite{s-fix}.

Within the global supersymmetry, with nonvanishing $F$-terms of Eq. (\ref{Fterms}), the corresponding potential would be
$V_F=(\lam_e^2+\lam_{\mu })s^2z^4$. This quartic potential (if $z$'s kinetic term is canonical) gives an unacceptably
 large tensor-to-scalar ratio and should be
refuted. However, within supergravity, the form of the  K\" ahler potential plays an essential role
\cite{Kallosh:2013hoa, Ferrara:2016fwe}, and we will consider the one given in Eq. (\ref{totalK}).
Below, the inflation within this simple setup is investigated.

\section{III. Inflation}
\vs{-0.3cm}

Before proceeding to study the inflationary scenario, we derive the inflaton potential.
Within $N=1$ SUGRA with  K\" ahler potential ${\cal K}$ and superpotential $W$, the $F$-term scalar potential
is given by \cite{sugra-pot}
\begin{eqnarray}
& V_F=\,e^{{\cal K}}\left(D_{\bar J}{\bar W}{\cal K}^{{\bar J}I}D_IW - 3|W|^2\right),
\label{sugrVF}
\end{eqnarray}
where $D_I=\partial_I+{\cal K}_I$ (with $\partial_I= \fr{\pl }{\pl \Phi_I}$, ${\cal K}_I= \fr{\pl {\cal K}}{\pl \Phi_I}$)
and we set the reduced Planck mass $M_{Pl}$ to one. Further, omitting $M_{Pl}$, any dimensionful quantity will be given in the unit of  $M_{Pl}$($=2.4\cdot 10^{18}$~GeV).
The matrix ${\cal K}^{{\bar J}I}$ is an inverse of the  matrix ${\cal K}_{I{\bar J}}=\fr{\pl^2 {\cal K}}{\pl \Phi_I \pl \Phi_I^\dag }$
(e.g., ${\cal K}_{I{\bar M}}{\cal K}^{{\bar M}J}=\de_I^J,~{\cal K}^{{\bar I}M}{\cal K}_{M{\bar J}}=\de^{\bar I}_{\bar J}$).

From Eqs. (\ref{WMSSM}) and (\ref{Yuk-basis}) with the VEV configuration (\ref{VEVs}), we have $\lan W\ran =0$, two nonvanishing
$F$-terms of Eq. (\ref{Fterms}), and thus
\begin{eqnarray}
&V_F=e^{\cal K}\l {\cal K}^{{e^{-}}^\dag e^{-}}|F_{e^{-}}|^2 +{\cal K}^{{\mu^c}^\dag \mu^c}|F_{\mu^c}|^2\r .
\la{V}
\end{eqnarray}
As already noted, for the K\" ahler potential we use the form (\ref{totalK}),  involving all MSSM states.
Canonical normalization of the inflaton field is dictated from
\begin{eqnarray}
&{\cal K}_{I\bar J}\pl \Phi_I \pl \Phi_J^{*}\to 3\fr{(\pl z)^2}{(1-3z^2)^2}\equiv \fr{1}{2}(\pl \phi)^2
\la{kin}
\end{eqnarray}
which gives the relation
\begin{eqnarray}
&z=\fr{1}{\sqrt{3}}\tanh (\fr{\phi }{\sqrt{2}}) ~,
\la{z-t}
\end{eqnarray}
where $\phi $ is the canonically normalized inflaton. With these results, from (\ref{V}) we derive the inflaton potential:
\begin{eqnarray}
&{\cal V}(\phi )=s^2\fr{\lam_{\mu}^2+\lam_e^2}{9}\tanh^4 (\fr{\phi }{\sqrt{2}}).
\la{Vinf}
\end{eqnarray}
Successful inflation is insured by the flat shape of the $\tanh \fr{\phi }{\sqrt{2}}$ function for large values of $\phi $.
The slow roll parameters derived from the potential - the "VSR" parameters - are given by
\cite{Stewart:1993bc, Liddle:1994dx}
\beq
\ep =\fr{1}{2}\l \fr{{\cal V}'}{{\cal V}}\r^2,~~
\eta =\fr{{\cal V}''}{{\cal V}},~~
\xi =\fr{{\cal V}'{\cal V}'''}{{\cal V}^2}.
\la{ep-eta}
\eeq
From these, the observable $n_s$ and $r$ will be determined. The parameters in (\ref{ep-eta}) are
independent of the $s$ [the factor appearing in (\ref{Vinf})]. The latter, as mentioned above, will be determined from
the observed value of $A_s$.

While the  parameters of (\ref{ep-eta}) are useful in the slow roll regime, for the determination of $\phi_e$ - the
point at which the inflation ends - we use the exact condition $\ep_H=1$ for the HSR parameter (derived from the Hubble parameter).
(For relations between HSR and VSR parameters see \cite{Stewart:1993bc}, \cite{Liddle:1994dx}.)
The point $\phi_i$ \cite{def-ie} - at the beggining of inflation - can be found from the demand of obtaining
a desirable value of the number
 of $e$-foldings $N_e^{\rm inf}$ during inflation, which is given via the HSR parameter $\ep_H$ by the exact expression
\begin{eqnarray}
&N_e^{\rm inf}=\fr{1}{\sqrt{2}}\int_{\phi_e}^{\phi_i} \fr{1}{\sqrt{\ep_H}} d\phi ~.
\la{exact-Ninf}
\end{eqnarray}
Given the value of $\phi_i$, we can calculate the spectral index $n_s$ and the tensor-to-scalar ratio $r$ by:
\begin{eqnarray}
&n_s&=1-6\ep_i+2\eta_i +\fr{2}{3}(22-9C)\ep_i^2-
\nonumber \\
& &-(14-4C)\ep_i \eta_i +\fr{2}{3}\eta_i^2+\fr{1}{6}(13-3C)\xi_i ~,
\nonumber \\
&r&=16\ep_i [ 1-(\fr{2}{3}-2C)(2\ep_i-\eta_i) ], ~~ C=0.0815.
\la{r-ns-SRV}
\end{eqnarray}
These expressions are valid within the second order approximation (which turns out to be pretty accurate because of the slow
roll regime at the $\phi_i$ point).

In order to guarantee the causality of fluctuations, the value of  the $N_e^{\rm inf}$
  should satisfy \cite{Liddle:2003as}:
\begin{eqnarray}
&N_e^{\rm inf}=&62-\ln \fr{k}{a_0H_0}-\ln \fr{10^{16}{\rm GeV}}{{\cal V}_i^{1/4}}+\ln \fr{{\cal V}_i^{1/4}}{{\cal V}_e^{1/4}}
 \nonumber \\
&&-\fr{4-3\ga }{3\ga }\ln \fr{{\cal V}_{e}^{1/4}}{\rho_{reh}^{1/4}}~,
\la{Ne}
\end{eqnarray}
where $k=0.002\hspace{0.6mm}{\rm Mpc}^{-1}$ (the present horizon scale is $a_0H_0\approx 0.00033\hspace{0.6mm}{\rm Mpc}^{-1}$).
The factor
$\ga =2\fr{\int_0^{\phi_e} (1-{\cal V}/{\cal V}_e)^{1/2}d\phi }{\int_0^{\phi_e} (1-{\cal V}/{\cal V}_e)^{-1/2}d\phi }$
(which turn out to be  $\simeq 1.19$)
 stands for the effect of the inflaton's oscillation  \cite{Turner:1983he} after inflation.

We have to reconcile the values of $N_e^{\rm inf}$ obtained from Eqs. (\ref{exact-Ninf}) and (\ref{Ne}).
 The values of ${\cal V}_i$ and ${\cal V}_e$, entering in Eq.  (\ref{Ne}), can be calculated with the help
of another observable - the amplitude of the curvature perturbation $A_s$, which, according to the Planck measurements
\cite{Akrami:2018odb}, should be $A_s^{1/2}= 4.581 \tm 10^{-5}$.
 Generated by inflation, this parameter is given by
\begin{eqnarray}
A_s^{1/2}=\fr{1}{\sqrt{12}\pi }\left | \fr{{\cal V}^{3/2}}{M_{Pl}^3{\cal V}'}\right |_{\phi_i}~.
\la{dT-Ts}
\end{eqnarray}
From this we can fix the numerical value of the parameter $s$ [see Eq. (\ref{Vinf}) where $s$ appears].

In addition, we need to calculate  $\rho_{\rm reh}\!\!=\!\fr{\pi^2}{30}g_*T_r^4$ through the reheating temperature, given by \cite{Kofman:1997yn}
\begin{eqnarray}
T_r=\l \fr{90}{\pi^2g_*}\r^{1/4}\!\!\sqrt{M_{Pl}\Ga(\phi )},
\la{r-reh1}
\end{eqnarray}
where $g_*$ is the effective  number of massless degrees at temperature $T_r$ and $\Ga(\phi )$ is the inflaton's decay width.
It turns, out that within our model, after determining the parameter $s$ (via the value of $A_s^{1/2})$ we will be able to calculate
$\Ga(\phi )$ and therefore predict the value of $T_r$.

Since the inflaton field comes from the MSSM fields, its couplings are also well fixed.
In particular, with canonically normalized fermion fields, the inflaton's interaction with quarks arises via the couplings
\begin{eqnarray}
&\fr{s}{\sqrt{3}}F(\phi )d^TY_Dd^c,~&
F(\phi )\!=\!\tanh \!\fr{\phi }{\sqrt{2}} (1\!-\!\tanh^2 \!\!\fr{\phi }{\sqrt{2}})^{1/2}.~~
\la{inf-ff}
\end{eqnarray}
Thus, for $\phi \!-\!d\!-\!d^c$ -type interactions  we obtain
\begin{eqnarray}
F'\fr{s}{\sqrt{3}}\phi d^TY_Dd^c ,~~~F'=\fr{dF}{d \phi} .
\la{inf-dd}
\end{eqnarray}
It turns out that the $\phi $ dominantly decays into a strange quark-antiquark pair.
From (\ref{inf-dd}) coupling we have
\begin{eqnarray}
\label{inf-width}
&\Ga (\phi )\!\simeq \!\Ga (\phi \to s\bar s)\!= \!
\fr{m_{\phi }}{16\pi }(s\lam_sF')^2\!\l 1-\fr{4m_s^2}{m_{\phi }^2}\r^{\!3/2}~,
 \nonumber \\
&m_s=\fr{s}{\sqrt{3}}F\lam_s, ~~m_{\phi }^2={\cal V}'' ,
\end{eqnarray}
where in (\ref{inf-width}) all $\phi $ dependent quantities are evaluated at $\phi =\phi_e$ (note that over the course
of inflation, the $\phi $ gets large values \cite{symmetries}).
With all of these results, we can carry out a detailed analysis of the inflation process and calculate all observable \cite{scale-RG}.
Our results are summarized as follows:
\begin{eqnarray}
\label{result-sum}
&n_s=0.9664 , ~~r=0.00117 ,~~\fr{dn_s}{d\ln k}=-5.94\cdot 10^{-4},
 \nonumber \\
&N_e^{\rm inf}=57.97 , ~~~\rho_{\rm reh}^{1/4}=1.58\cdot 10^{8}{\rm GeV},
\nonumber \\
&T_r=7.22\cdot 10^{7}{\rm GeV},
\end{eqnarray}
where the value of the spectral index running (given by
$ \fr{dn_s}{d\ln k}=16\ep_i\eta_i-24\ep_i^2-2\xi_i$) is also included.
These results are rather insensitive to the values of $\tan \bt $($=v_u/v_d$, determining the strength of the Yukawa couplings), because the combination $s\lam_{\mu }$ (and also
$s\lam_s$ which enters in the inflaton decay width) is fixed from the value of $A_s^{1/2}$ by
proper selection of the factor $s$.
The values of $n_s, r$ and $\fr{dn_s}{d\ln k}$, given in the first line of Eq. (\ref{result-sum}), are in good agreement
with the current observations \cite{Akrami:2018odb}.
The obtained value of $T_r$ [see Eq. (\ref{result-sum})], for specific and phenomenologically viable sparticle spectroscopy,
can be compatible with the upper bounds \cite{TR-bounds} needed to avoid the relic gravitino problem.

\section{IV. Summary and Outlook}
\vs{-0.3cm}

In this paper, aiming for minimalistic construction, we have presented a novel inflationary scenario
within the MSSM (without any extension). Within the model, the inflaton field is a combination
of slepton and down-type Higgs doublet states whose VEVs are aligned toward the flat $D$-term direction.
Nonzero vacuum energy, driving the inflation, is due to the MSSM leptonic Yukawa superpotential terms.
In particular, inflation is mainly due to the muonic Yukawa coupling. At the same time, the
reheating process is via  inflaton $\phi \to s\bar s$ decay, controlled by the strange quark Yukawa interaction.
These results, together with the successful inflationary scenario, allow for having the predictive scheme.

Since the inflaton field involves the VEVs of the slepton states, the lepton number is violated and, as
 an additional outcome of the model, it would be interesting to investigate the impact of this effect on the leptogenesis
 process in the spirit of Refs. \cite{Affleck:1984fy, Dine:1995kz}. Moreover, it would be interesting to investigate if an analogous
 (with only MSSM Yukawa couplings, along the $D$-flat direction) inflationary scenario would work with the
 inflaton  involving the squark VEVs.

Finally, it would be highly motivated, although quite challenging, to realize ideas developed in this work
within the grand unified models based on symmetries such as $SU(5)$, $SO(10)$, etc.

We reserve these exciting issues for future investigation and also hope that the presented work will
motivate others' research along these lines.

\begin{acknowledgments}
\end{acknowledgments}
%

%


\begin{thebibliography}{26}
\bibitem{s-fix}
One could also consider alternative ways/mechanisms for fixing $s$ and investigate various phenomenological implications.
These issues are left for  future studies.
\bibitem{scale-RG}
Diring the calculations, since inflation occurs at high scales, for convenience we have RG
evaluated masses and couplings at the scale $M_{Pl}$.
\end{thebibliography}
\end{document}